\documentclass[a4paper]{article}

\usepackage{pict2e}
\usepackage{amssymb,amsmath,array}
\usepackage[dvips]{graphicx,epsfig}
\usepackage[usenames]{color}
\usepackage{wrapfig,rotating}
\usepackage{cite}
 
\newcommand{\red}{\color{red}}%
\newcommand{\green}{\color{green}}%
\newcommand{\blue}{\color{blue}}%
\newcommand{\yellow}{\color{yellow}}%
\definecolor{goldenrod}{named}{Goldenrod}%
\definecolor{dandelion}{named}{Dandelion}%
                                         \newcommand{\lorange}{\color{dandelion}}%
\definecolor{pinegreen}{named}{PineGreen}%
\definecolor{limegreen}{named}{LimeGreen}%
\definecolor{bluegreen}{named}{BlueGreen}%
                                         \newcommand{\bgreen}{\color{bluegreen}}%
\definecolor{olivegreen}{named}{OliveGreen}%
                                         \newcommand{\ogreen}{\color{olivegreen}}%
\definecolor{aquamarine}{named}{Aquamarine}\newcommand{\aquamarine}{\color{aquamarine}}%
\definecolor{cornflowerblue}{named}{CornflowerBlue}%
\definecolor{royalblue}{named}{RoyalBlue}%
                                         \newcommand{\rblue}{\color{royalblue}}%
\definecolor{midnightblue}{named}{MidnightBlue}%
\definecolor{violet}{named}{Mulberry}\newcommand{\violet}{\color{violet}}%
\definecolor{redviolet}{named}{RedViolet}%
\definecolor{wildstrawberry}{named}{WildStrawberry}%
                                                   \newcommand{\dred}{\color{wildstrawberry}}%
\definecolor{purple}{named}{Purple}\newcommand{\purple}{\color{purple}}%
\definecolor{orange}{named}{Orange}%
\definecolor{burntorange}{named}{BurntOrange}\newcommand{\burntorange}{\color{burntorange}}%
\definecolor{rawsienna}{named}{RawSienna}%
                                         \newcommand{\lbrown}{\color{rawsienna}}%
\definecolor{brown}{named}{Brown}\newcommand{\brown}{\color{brown}}%
\definecolor{sepia}{named}{Sepia}%
                                 \newcommand{\dbrown}{\color{sepia}}%
\begin{document}
\begin{center}
{\Large Bose-Einstein Correlations and the Tau Model}\footnote{To appear in proceedings
\textit{XL International Symposium on Multiparticle Dynamics, Antwerp, Sept.\ 2010}} \\[1.1ex]
 
{          W.J. Metzger\/$^1$\footnote{Speaker},
                  T. Nov\'ak\/$^2$,
                  T. Cs\"org\H{o}\/$^3$,
                  W. Kittel\/$^1$ \\ for the L3 Collaboration\\[1ex]
$^1$Radboud University, P.O. Box 9044, 6500 KD\ \ Nijmegen, Netherlands\\
$^2$K\'aroly R\'obert College, 3200 Gy\"ongy\"os, Hungary\\
$^3$MTA KFKI RMKI, 1525 Budapest 114, Hungary, and Harvard University,\\ 17 Oxford St., Cambridge, MA 02138,
U.S.A.}
\end{center}


\newcommand{\taumodel}{$\tau$-model}
\newcommand{\adhoc}{\textit{ad hoc}}
\newcommand{\apriori}{{\it a priori}}
\newcommand{\ca}{{\it ca.}}%
\newcommand{\cf}{{\it cf.}}%
\newcommand{\etal}{{\it et~al.}}%
\newcommand{\etc}{{\it etc.}}%
\newcommand{\eg}{{\it e.g.}}%
\newcommand{\ie}{{\it i.e.}}%
\newcommand{\vs}{{\it vs.}}%
\newcommand{\Eq}[1]{Eq.\,(\ref{#1})}%
\newcommand{\Eqs}[1]{Eqs.\,(\ref{#1})}%
\newcommand{\Fig}[1]{Fig.\,\ref{#1}}%
\newcommand{\Figs}[1]{Figs.\,\ref{#1}}%
\newcommand{\Tab}[1]{Table~\ref{#1}}%
\newcommand{\Tabs}[1]{Tables~\ref{#1}}%
\newcommand{\Lthree}{{\scshape l}{\small 3}}
\newcommand{\Pep}{e$^+$}%
\newcommand{\Pem}{e$^-$}%
\newcommand{\Pgp}{\ensuremath{\pi}}%
\newcommand{\PZ}{\ensuremath{\mathrm{Z}}}
\newcommand{\eV}{\hbox{\ensuremath{\mathrm{e\kern-0.1em V}}}}%
\newcommand{\MeV}{\hbox{\ensuremath{\mathrm{M}}\eV}}%
\newcommand{\GeV}{\hbox{\ensuremath{\mathrm{G}}\eV}}%
\newcommand{\invGeV}{\GeV\ensuremath{^{-1}}}
\newcommand{\abs}[1]{\left|#1\right|}                        
\newcommand{\chisq}{\ensuremath{\chi^2}}
\newcommand{\ycut}{\ensuremath{y_\mathrm{cut}}}
\newcommand{\kt}{\ensuremath{k_\mathrm{t}}}
\newcommand{\pt}{\ensuremath{p_\mathrm{t}}}
\newcommand{\mt}{\ensuremath{m_\mathrm{t}}}

\newcommand{\Qsquare}{\ensuremath{Q^2}}
\newcommand{\Qtrans}{\ensuremath{Q_\mathrm{T}}}
\newcommand{\Qlong}{\ensuremath{Q_\mathrm{L}}}
\newcommand{\Qside}{\ensuremath{Q_\mathrm{side}}}
\newcommand{\Qout}{\ensuremath{Q_\mathrm{out}}}
\newcommand{\qout}{\ensuremath{q_\mathrm{out}}}
\newcommand{\Qle}{\ensuremath{Q_\mathrm{LE}}}
\newcommand{\Qstrans}{\ensuremath{Q^2_\mathrm{T}}}
\newcommand{\Qslong}{\ensuremath{Q^2_\mathrm{L}}}
\newcommand{\Qsside}{\ensuremath{Q^2_\mathrm{side}}}
\newcommand{\Qsout}{\ensuremath{Q^2_\mathrm{out}}}
\newcommand{\qsout}{\ensuremath{q^2_\mathrm{out}}}
\newcommand{\Qsle}{\ensuremath{Q^2_\mathrm{LE}}}
\newcommand{\Rsquare}{\ensuremath{R^2}}
\newcommand{\Rtrans}{\ensuremath{R_\mathrm{T}}}
\newcommand{\Rlong}{\ensuremath{R_\mathrm{L}}}
\newcommand{\Rside}{\ensuremath{R_\mathrm{side}}}
\newcommand{\Rout}{\ensuremath{R_\mathrm{out}}}
\newcommand{\Ralong}{\ensuremath{R_{\mathrm{a}\;\mathrm{L}}}}
\newcommand{\Raside}{\ensuremath{R_{\mathrm{a}\;\mathrm{side}}}}
\newcommand{\Raout}{\ensuremath{R_{\mathrm{a}\;\mathrm{out}}}}
\newcommand{\rhoout}{\ensuremath{\rho_\mathrm{out}}}
\newcommand{\rout}{\ensuremath{r_\mathrm{out}}}
\newcommand{\Rstrans}{\ensuremath{R^2_\mathrm{T}}}
\newcommand{\Rslong}{\ensuremath{R^2_\mathrm{L}}}
\newcommand{\Rsside}{\ensuremath{R^2_\mathrm{side}}}
\newcommand{\Rsout}{\ensuremath{R^2_\mathrm{out}}}
\newcommand{\rhosout}{\ensuremath{\rho^2_\mathrm{out}}}
\newcommand{\rsout}{\ensuremath{r^2_\mathrm{out}}}
\newcommand{\Rle}{\ensuremath{R_\mathrm{LE}}}
\newcommand{\Rsle}{\ensuremath{R^2_\mathrm{LE}}}
\newcommand{\epsiltrans}{\ensuremath{\epsilon_\mathrm{T}}}
\newcommand{\epsillong}{\ensuremath{\epsilon_\mathrm{L}}}
\newcommand{\epsilside}{\ensuremath{\epsilon_\mathrm{side}}}
\newcommand{\epsilout}{\ensuremath{\epsilon_\mathrm{out}}}
\newcommand{\epsille}{\ensuremath{\epsilon_\mathrm{LE}}}
\newcommand{\Deltatautrans}{\ensuremath{\Delta\tau_\mathrm{T}}}
\newcommand{\Deltataulong}{\ensuremath{\Delta\tau_\mathrm{L}}}
\newcommand{\Deltatauside}{\ensuremath{\Delta\tau_\mathrm{side}}}
\newcommand{\Deltatauout}{\ensuremath{\Delta\tau_\mathrm{out}}}
 
\newcommand{\Ra}{\ensuremath{R_\mathrm{a}}}
\newcommand{\pho}{\phantom{0}}

\begin{abstract}
Bose-Einstein correlations of pairs of identical charged pions
produced in hadronic Z decays are analyzed in terms of various parametrizations.
A good description is achieved using a L\'evy stable distribution in conjunction with a
model where a particle's momentum is highly correlated with its space-time point of
production, the \taumodel.
However, a small but significant elongation of the particle emission region is observed
in the Longitudinal Center of Mass frame, which is not accommodated in the \taumodel.
This is investigated using an \adhoc\/ modification of the \taumodel.
\end{abstract}

\section{Introduction}\label{sect:intr}
We study Bose-Einstein correlations (BEC) in hadronic \PZ\ decay
using data collected by the
\Lthree\ detector 
at an \Pep\Pem\    center-of-mass energy of $\sqrt{s}\simeq 91.2$ \GeV.
Approximately 36 million like-sign pairs of well-measured charged tracks from about 0.8 million
hadronic Z decays are used~\cite{tamas:thesis}.
Events are classified as two- or three-jet events
using calorimeter clusters with the Durham jet
algorithm  
with a jet resolution parameter $\ycut=0.006$, yielding about 0.5 million two-jet events.
There are few events with more than three jets, and they are included in the three-jet sample.
To determine the thrust axis of the event we also use calorimeter clusters.

The two-particle correlation function of two particles with
four-momenta $p_{1}$ and $p_{2}$ is given by the ratio of the two-particle number density,
$\rho_2(p_{1},p_{2})$,
to the product of the two single-particle 
densities, $\rho_1 (p_{1})\rho_1 (p_{2})$.
Since we are 
interested only in BEC,
$\rho_1 (p_{1})\rho_1 (p_{2})$  is replaced by $\rho_0(p_1,p_2)$,
the two-particle density that would occur in the absence of BEC, yielding $  R_2(p_1,p_2)=     {\rho_2(p_1,p_2)}/{\rho_0(p_1,p_2)}$.
An event mixing technique is used to construct $\rho_0$.
 
Since the mass of the identical particles of the pair is fixed,
$R_2$
is defined in six-dimensional momentum space, which is often reduced to a single dimension,
$Q=\sqrt{-(p_1-p_2)^2}$.
 But there is no reason
to expect the hadron source to be spherically symmetric in jet fragmentation.
In fact, the source is found to be elongated along the
jet axis~\cite{L3_3D:1999,OPAL3D:2000,DELPHI2D:2000,ALEPH:2004},
                                             but only by                         about 25\%,
which suggests that a parametrization in terms of the single variable $Q$,
may be a good approximation.
This is also suggested by studies of various decompositions of $Q$~\cite{tamas:thesis,TASSO:1986}.

\section{Parametrizations of BEC}              \label{sect:param}
With a few assumptions, 
$R_2$
is related to the Fourier transform, $\tilde{f}(Q)$, of
the (configuration space) density distribution of the source, $f(x)$:
\begin{equation}  \label{eq:R2fourier}
     R_2(p_1,p_2) = \gamma \left[ 1 + \lambda |\tilde{f}(Q)|^2 \right]
                    \left(1 + \delta Q \right) \;.
\end{equation}
The parameter $\gamma$ and
the $(1 + \delta Q)$ term are
introduced to parametrize
possible long-range correlations not adequately accounted for in $\rho_0$, 
and $\lambda$ to account for several factors, such as lack of complete
incoherence of particle production and presence of long-lived resonance decays.

Model-independent ways to study deviations from
the Gaussian
are to expand  $\tilde{f}(Q)$ 
about a Gaussian using 
the Edgeworth expansion 
or to replace $f(Q)$ by a symmetric L\'evy distribution.
%

Both the symmetric L\'evy
(Fig.~1a)
and the Edgeworth parametrizations do a fair job~\cite{wes:WPCF2006,wes:Hangzhou2006}
of describing the
region $Q<0.6\,\GeV$, but fail at higher $Q$, particularly the region 0.6--1.5\,\GeV\
where $R_2$ dips below unity,
indicative of an anti-correlation.
This is clearly seen in
Fig.~1a.

\begin{figure}[h]
  \centering
  \includegraphics[width=.32\textwidth,bb=58 87 519 682,clip]{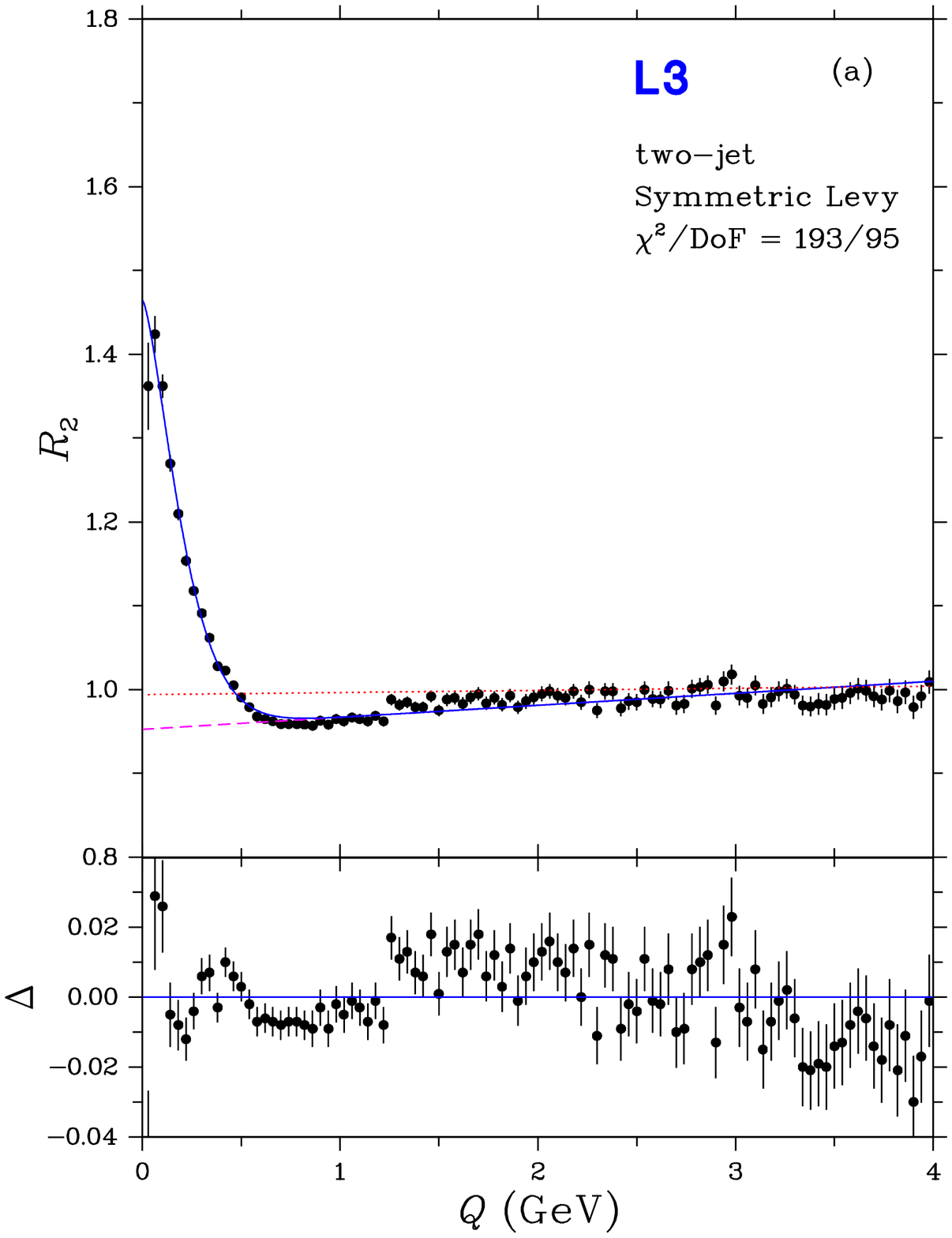}
 \hfil
  \includegraphics[width=.32\textwidth,bb=58 87 519 682,clip]{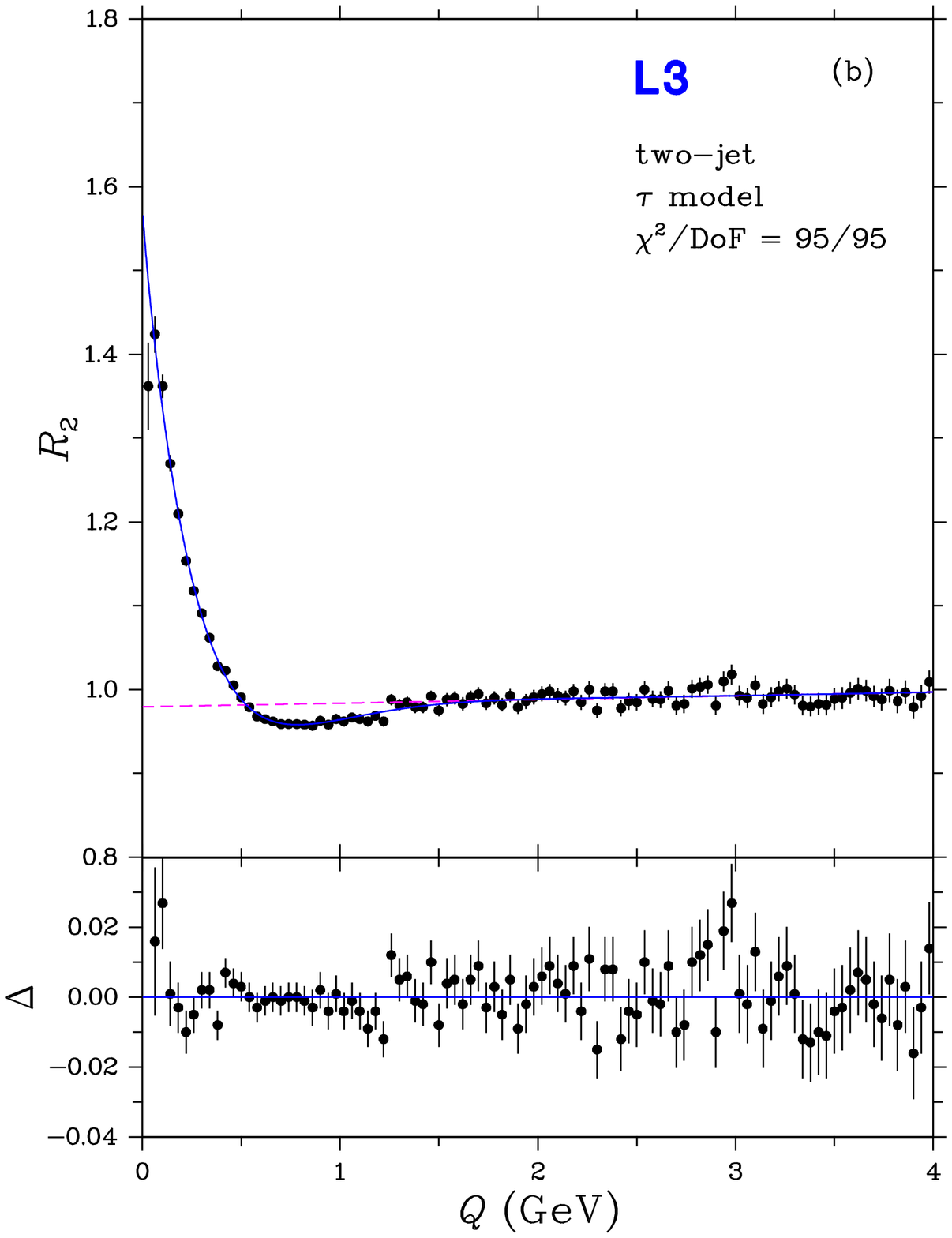}
 \hfil
  \includegraphics[width=.32\textwidth,bb=58 87 519 682,clip]{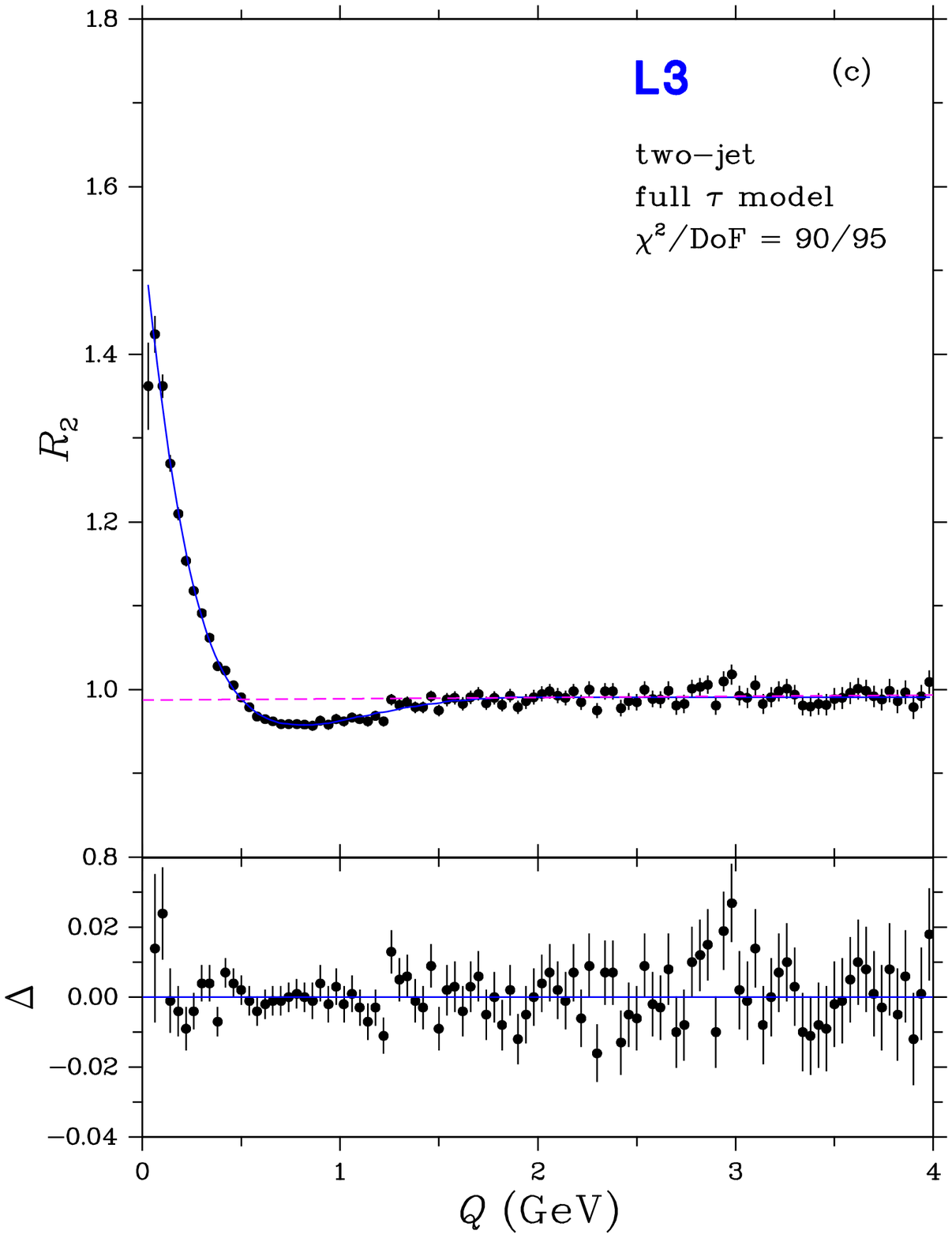}
  \caption{The Bose-Einstein correlation function $R_2$ for two-jet events. The curves
           correspond  to fits of the parametrizations: (a) the symmetric L\'evy;
           (b) the simplified \taumodel, \Eq{eq:asymlevR2};
           and (c) the full \taumodel, \Eq{eq:levyR2a}.
           Also plotted is $\Delta$, the difference between the fit and the data.
           The dashed line represents the long-range part of the fit, \ie, $\gamma(1+\epsilon Q)$.
           The dotted line in (a) represents a linear fit in the region $Q\ge1.5\,\GeV$.
           \label{fig:a_levy_2jet}
           }
\end{figure}

We
have seen that BEC depend, at least approximately, only on $Q$, 
not on its components separately and
%
that $R_2$ in the region 0.6--1.5\,\GeV\ dips below its values at higher $Q$.
A model which predicts such $Q$-dependence as well as
a dependence on the transverse mass,  $\mt=\sqrt{m^2+\pt^2}$, which has been previously
observed~\cite{Smirnova:Nijm96,Dalen:Maha98,OPAL:2007},
is the
$\tau$-model~\cite{Tamas;Zimanji:1990},
in which
it is assumed that
in the overall center-of-mass system
the average production point
$\overline{x}=(\overline{t},\overline{r}_x,\overline{r}_y,\overline{r}_z)$, of particles with a given
four-momentum $p$ is given by
$   \overline{x}^\mu (p^\mu)  = a\tau p^\mu$. 
In the case of two-jet events, $a=1/\mt$,
and
$\tau = \sqrt{\overline{t}^2 - \overline{r}_z^2}$ is the longitudinal proper time; 
for the case of three-jet events the relation is more complicated.
The second assumption is that the distribution of $x^\mu (p^\mu)$ about its average
is narrower than the proper-time distribution,   $H(\tau)$.
Then $R_2$ is found~\cite{ourTauModel}
to depend only on $Q$, the values of $a$ of the two pions, and
the Fourier transform of $H(\tau)$.
Since there is no particle production before the onset of the collision,
$H(\tau)$ should be a  one-sided distribution.
We choose a one-sided L\'evy distribution, which is characterized by three parameters:
the index of stability $\alpha$, the proper time of the start of particle emission $\tau_0$,
and $\Delta\tau$, which is a measure of the width of $H(\tau)$.
Then~\cite{ourTauModel}
\begin{equation} \label{eq:levyR2a}
\begin{split}
   R_2(Q,a_1,a_2) &= {} \gamma \left\{  1 +
      \lambda\cos\left[\frac{\tau_0 Q^2 (a_1+a_2)}{2} +
\tan\left(\frac{\alpha\pi}{2}\right)\left(\frac{\Delta\tau {Q^2}}{2}\right)^{\!\alpha}\frac{a_1^\alpha+a_2^\alpha}{2} \right]
  \right.
\\ 
     &
  \left.
     \quad \cdot           \exp \left[-\left(\frac{\Delta\tau {Q^2}}{2}\right)^{\!\alpha}\frac{a_1^\alpha+a_2^\alpha}{2} \right]
       \right\} \left(1+\epsilon Q\right)
 \; .
\end{split}
\end{equation}
 
Note that the cosine factor generates oscillations corresponding to alternating correlated and anti-correlated regions,
a feature clearly seen in the data (Fig.~1).
Note also that since $a=1/\mt$ for two-jet events, the \taumodel\ predicts a decreasing effective source size with
increasing \mt.
 
Before proceeding to fits of \Eq{eq:levyR2a},
we first consider a simplification of the equation
obtained by assuming (a) that particle production starts immediately, \ie, $\tau_0=0$,
and (b) an average $a$-dependence, which is implemented  
by introducing an effective radius, $R$, defined by
\begin{equation}\label{eq:effR}
    R^{2\alpha} = \left(\frac{\Delta\tau}{2}\right)^{\!\alpha} \frac{a_1^\alpha+a_2^\alpha}{2} \;.
\end{equation}
This results in
\begin{equation}\label{eq:asymlevR2}
    R_2(Q) = \gamma \left[ 1+ \lambda \cos \left(\left(R_\mathrm{a}Q\right)^{2\alpha} \right)
             \exp \left(-\left(RQ\right)^{2\alpha} \right) \right] (1+ \epsilon Q) \;,
\end{equation}
where $R_\mathrm{a}$ is  related to $R$ by
\begin{equation}\label{eq:asymlevRaR}
    R_\mathrm{a}^{2\alpha} = \tan\left(\frac{\alpha\pi}{2}\right) R^{2\alpha} \;.
\end{equation}
Fits of
\Eq{eq:asymlevR2} are first performed with $R_\mathrm{a}$ as a free parameter.
The fits for both two- and three-jet events
have acceptable confidence levels (CL), and describe well the dip 
in the 0.6--1.5\,\GeV\ region,
as well as the peak at low values of $Q$.
The estimates of some fit parameters are rather highly correlated.
For example, for two-jet events the estimated correlation coefficients from the fit for
$\alpha$, $R$ and $\Ra$ are $\rho(\alpha,R)=-0.62$, $\rho(\alpha,\Ra)=-0.92$, and $\rho(R,\Ra)=0.38$.
Taking the correlations into account, the fit parameters 
satisfy \Eq{eq:asymlevRaR},
the difference between the left- and right-hand sides of the equation being less than 1 standard deviation
for two-jet events and about 1.5 standard deviations for three-jet events.
%
No significant long-range correlation is observed: $\epsilon$ is zero within 1 standard deviation,
and fits with $\epsilon$ fixed to zero find the same values, within 1 standard deviation, of the parameters
as the fits with $\epsilon$ free.
Thus the method to remove non-Bose-Einstein correlations in the reference sample is apparently successful.
 
Fits are also performed imposing \Eq{eq:asymlevRaR}.
The result for two-jet events is shown in Fig.~1b.
For two-jet events, the values of the parameters are comparable to those with $\Ra$ free.
For three-jet events, the imposition of \Eq{eq:asymlevRaR} results in values of $\alpha$ and $R$
closer to those for two-jet events, but the $\chi^2$ is noticeably worse, though acceptable, than with $\Ra$ free.
Also,
the values of $\epsilon$ differ somewhat from zero, which could indicate a
slight deficiency in the description of BEC
similar to, but on a much smaller scale, the
Edgeworth and symmetric L\'evy fits.


For two-jet events, $a=1/\mt$, while for three-jet events the situation is more complicated.
We therefore limit fits of  \Eq{eq:levyR2a} to the two-jet data.
For each bin in $Q$ the average values of $m_{\mathrm{t}1}$ and $m_{\mathrm{t}2}$ are calculated,
where $m_{\mathrm{t}1}$ and $m_{\mathrm{t}2}$ are the transverse masses of the
two particles making up a pair, requiring  $m_{\mathrm{t}1} > m_{\mathrm{t}2}$.
Using these averages,
 \Eq{eq:levyR2a} is fit to $R_2(Q)$.
The fit yields  $\tau_0=0.00\pm0.02$~fm, and we have re-fit with $\tau_0$ fixed to zero.
The results are shown in Fig.~1c.
The parameters $\alpha$, $\Delta\tau$ and $\lambda$ are highly correlated with estimated correlaton coefficients
$\rho(\alpha,\Delta\tau)=-0.96$, $\rho(\alpha,\lambda)=-0.92$, and $\rho(\Delta\tau,\lambda)=0.95$.

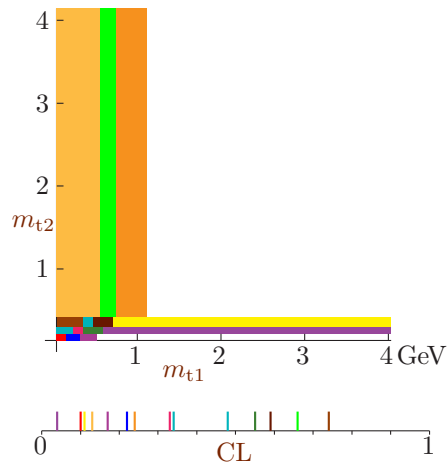
\begin{wrapfigure}{r}{0.4 \textwidth}
\begin{minipage}{.3\textwidth}
  \setlength{\unitlength}{0.11mm}
\vspace{-7mm}
  \begin{picture}(454,444)(-50,0)
     \put(0,14){\line(1,0){414}}
     \put(14,0){\line(0,1){414}}
    {\rblue
     \moveto(0,0) {\red\polygon*(14,14)(14,22)(26,22)(26,14)}             
     \moveto(0,0) {\aquamarine\polygon*(14,22)(14,30)(34,30)(34,22)}      
     \put(0,0){{\lbrown\polygon*(14,30)(46,30)(46,42)(14,42)}}            
     \moveto(0,0) {\lorange\polygon*(14,42)(66,42)(66,414)(14,414)}       
     \moveto(0,0) {\blue\polygon*(26,14)(26,22)(42,22)(42,14)}            
     \moveto(0,0) {\dred\polygon*(34,22)(46,22)(46,30)(34,30)}            
     \moveto(0,0) {\bgreen\polygon*(46,30)(58,30)(58,42)(46,42)}          
     \moveto(0,0) {\green\polygon*(66,42)(86,42)(86,414)(66,414)}        
     \moveto(0,0) {\violet\polygon*(42,14)(62,14)(62,22)(42,22)}          
     \moveto(0,0) {\ogreen\polygon*(46,22)(70,22)(70,30)(46,30)}          
     \moveto(0,0) {\dbrown\polygon*(58,30)(82,30)(82,42)(58,42)}          
     \moveto(0,0) {\burntorange\polygon*(86,42)(122,42)(122,414)(86,414)} 
     \moveto(0,0) {\purple\polygon*(70,22)(414,22)(414,30)(70,30)}        
     \moveto(0,0) {\yellow\polygon*(82,30)(82,42)(414,42)(414,30)}        
    }
     \multiput(100,14)(100,0){4}{\line(0,1){8}}
     \put(-21,90){1}
     \put(-21,190){2}
     \put(-50,150){{\brown$m_{{\mathrm t}2}$}}
     \put(-21,290){3}
     \put(-21,390){4}
     \put(90,-15){1}
     \put(190,-15){2}
     \put(130,-30){{\brown$m_{{\mathrm t}1}$}}
     \put(290,-15){3}
     \put(390,-15){4}
     \put(410,-15){\GeV}
     \multiput(4,100)(0,100){4}{\line(1,0){8}}
  \end{picture}
\end{minipage}
\\[5mm]
\begin{minipage}{.39\textwidth}
  \setlength{\unitlength}{0.51mm}
  \begin{picture}(110,20)(-10,-10)
     \put(0,0){\line(1,0){100}}
     \multiput(0,0)(10,0){11}{\line(0,-1){1}}
     \put(-2,-6){0}
     \put(98,-6){1}
     \put(45,-8){{\brown CL}}
   \thicklines
     \put(10,0){\red        \line(0,1){5}}
     \put(48,0){\aquamarine \line(0,1){5}}
     \put(74,0){\lbrown     \line(0,1){5}}
     \put(13,0){\lorange    \line(0,1){5}}
     \put(22,0){\blue       \line(0,1){5}}
     \put(33,0){\dred       \line(0,1){5}}
     \put(34,0){\bgreen     \line(0,1){5}}
     \put(66,0){\green      \line(0,1){5}}
     \put(17,0){\violet     \line(0,1){5}}
     \put(55,0){\ogreen     \line(0,1){5}}
     \put(59,0){\dbrown     \line(0,1){5}}
     \put(24,0){\burntorange\line(0,1){5}}
     \put( 4,0){\purple     \line(0,1){5}}
     \put(11,0){\yellow     \line(0,1){5}}
  \end{picture}
\end{minipage}
  \caption{The regions of the $m_{\mathrm{t}1}$-$m_{\mathrm{t}2}$ plane
           and the CLs of the fits in these regions.
           \label{fig:mtmt}
           }
\vspace{-4mm}
\end{wrapfigure}

Since the \taumodel\ describes the \mt\ dependence of $R_2$,
                                        its parameters, $\alpha$,  $\Delta\tau$, and $\tau_0$, should not depend on \mt.
However, $\lambda$, which is not a parameter of the \taumodel,
but rather a measure of the strength of the BEC, can depend on \mt.
The large correlation between the fit estimates of $\lambda$, $\alpha$, and  $\Delta\tau$
complicate the testing of \mt-independence.  We perform fits in various regions of the $m_{\mathrm{t}1}$-$m_{\mathrm{t}2}$ plane
keeping $\alpha$ and $\Delta\tau$ fixed at the values obtained in the fit to the entire \mt\ plane.
The regions are chosen such that the numbers of pairs of particles in the regions are comparable.
The \mt\ regions and the CLs of the fits are shown in \Fig{fig:mtmt}.
The CLs are reasonably uniformly distributed between 0 and 1.
The data are thus in agreement with the hypothesis of \mt-independence of the parameters of the \taumodel.
 
\section{Do               BEC depend on components of \boldmath{$Q$}?}            \label{sect:elongation}
 
In this section we return to the question of a possible elongation of the region of homogeneity,
as viewed in the rest frame of the pion pair, and the possible dependence of the BEC correlation
function on components of $Q$.
 
The \taumodel\ predicts that the two-particle BEC correlation function $R_2$ depends on the two-particle
momentum difference only through $Q$, not through components of $Q$. However,
previous studies~\cite{L3_3D:1999,OPAL3D:2000,DELPHI2D:2000,ALEPH:2004,OPAL:2007} have found
that $R_2$ does depend on components of $Q$ and indeed that the shape of the region of homogeneity is
elongated along the event (thrust) axis.
The question is whether this is
an artifact of the Edgeworth or Gaussian parametrizations used in these studies
or shows a defect of the \taumodel.
 
In the studies of elongation, $Q^2$ is split into three components
in the Longitudinal Center of Mass System (LCMS) of the pair:
\begin{align}   \label{eq-QLCMS}
  \Qsquare &=  \Qslong + \Qsside + \Qsout - (\Delta E)^2   \\
           &=  \Qslong + \Qsside + \Qsout \left(1 - \beta^2\right)  \;,  \qquad
 \text{where}\quad \beta   = \frac{p_{1\mathrm{out}}+p_{2\mathrm{out}}}{E_1+E_2} \;.
\end{align}
The LCMS frame is defined as the frame, obtained by a Lorentz boost along the event axis,
where the sum of the three-momenta of the two pions ($\vec{p}_1+\vec{p}_2$)
is perpendicular to the event axis.
Assuming azimuthal symmetry about the event axis suggests that the region of
homogeneity have an ellipsoidal  shape with one axis, the longitudinal axis, along the event axis.
 
In the LCMS frame the event axis is referred to as the longitudinal direction,
the direction of $\vec{p}_1+\vec{p}_2$ as the out direction, and
the direction perpendicular to these as the side direction.
The components of $\abs{\vec{p}_1-\vec{p}_2}$ along these directions are denoted
\Qlong, \Qout, and \Qside, respectively.
$\Rsquare\Qsquare$ is then replaced in the Gaussian or Edgeworth parametrizations by
\begin{equation}   \label{eq-RQLCMS}
    \Rsquare\Qsquare  \Longrightarrow \Rslong\Qslong + \Rsside\Qsside + \rhosout\Qsout \;.
\end{equation}
\Rlong\ and \Rside\ are measures of the longitudinal and transverse size of the source, respectively,
whereas the value of \rhoout\ reflects both the transverse and temporal sizes.\footnote{In the
literature the coefficient of \Qsout\ in \Eq{eq-RQLCMS} is usually denoted \Rsout. We prefer to use
\rhosout\ to emphasize that, unlike \Rlong\ and \Rside, \rhoout\ contains a dependence on $\beta$.
}
 
Since the present analysis uses a somewhat different event sample and mixing algorithm, we have first
checked that we
find values of $\Rside/\Rlong$ consistent with our previous analysis~\cite{L3_3D:1999}.

We next investigate whether the \taumodel\ parametrizations could accommodate an elongation
by incorporating \Eq{eq-RQLCMS}
in \Eq{eq:asymlevR2}.  If \Eq{eq:asymlevRaR} is imposed, this results in
\begin{equation}\label{eq-asymlevR2BP}
    R_2(Q) = \gamma \left[ 1+ \lambda
                       \cos\left(\!\tan\left(\frac{\alpha\pi}{2}\right) A^{2\alpha} \vphantom{\frac{\Gamma}{2}}
                          \right)
             \exp \left(-A^{2\alpha} \right) \right]
             \left(1+ \epsillong\Qlong + \epsilside\Qside + \epsilout\Qout\right) \;,
\end{equation}
where
\begin{equation}\label{eq-asymlevR2BPA}
        A^2 = \Rslong\Qslong + \Rsside\Qsside + \rhosout\Qsout  \;.
\end{equation}
A fit of \Eq{eq-asymlevR2BP} for two-jet events
for $Q<4\,\GeV$ has a CL of 66\%, and
a clear preference for elongation is found: $\Rside/\Rlong=0.61\pm0.02$.
The value of $R$ found in the \taumodel\ fit shown in Fig.~1c
lies between the values of $\Rside$ and $\Rlong$ found here.
Furthermore, the value of $\Rside/\Rlong$ agrees well with that found using the Edgeworth parametrization.

%
To directly test the hypothesis of no separate dependence on components of $Q$, we investigate
two decompositions of $Q$:
\begin{subequations}
\begin{align}               
  \Qsquare &=  \Qsle   + \Qsside + \Qsout \;,& \Qsle &=\Qslong - (\Delta E)^2\;;  \label{eq-Q2dec-el} \\
  \Qsquare &=  \Qslong + \Qsside + \qsout \;,& \qsout &=\Qsout - (\Delta E)^2\;.  \label{eq-Q2dec-q}
\end{align}
\end{subequations}
The first, \Eq{eq-Q2dec-el}, corresponds to the LCMS frame where the longitudinal and energy terms
are combined; its three components of $Q$ are invariant with respect to Lorentz boosts along the event
axis. The second, \Eq{eq-Q2dec-q}, corresponds to the LCMS frame boosted to the rest frame of the pair;
its three components are invariant under Lorentz boosts along the out direction.
 
The test then consists of replacing \Rsquare\Qsquare\ by \Rsquare\ times one of the above equations and
comparing a fit where the coefficients of all three terms are constrained to be equal to a fit where each
coefficient is a free parameter.
For both parametrizations the fit allowing separate dependence on the components of $Q$ has a significantly better
CL
than the fit which does not.
The fit using
\Eq{eq-Q2dec-el}
attains a CL 
of 2\% when no separate dependence is allowed.
While this is not poor enough to reject, by itself,  the hypothesis of separate dependence,
the fit allowing it is, with 
CL = 38\%, much better.
When
\Eq{eq-Q2dec-q}
is used the fit not allowing separate dependence is rejected by a CL 
of $10^{-7}$, while allowing it results in the
marginally acceptable CL 
of 2\%.
For both decompositions
the differences in \chisq\ between
the fit allowing separate dependence
and that not allowing it is huge, 296 and 464, respectively, while a difference of only 2
is expected if there is no separate dependence.
This provides very strong evidence in favor of separate dependence on the components of $Q$.
Moreover, the radii corresponding to \Qside\ and \qout\ are found to be unequal by about a factor 2,
which violates the assumption of azimuthal symmetry assumed in the LCMS analysis.

\section{Discussion}  \label{sect:discussion}
The usual parametrizations of BEC of pion pairs in terms of $Q$,
such as the Gaussian,  Edgeworth, or symmetric L\'evy parametrizations, are found to be incapable
of describing the BEC correlation function $R_2$, particularly in the anti-correlation region,
$0.5<Q<1.5$\,\GeV.

The \taumodel\ assumes a very high degree of correlation between momentum space and space-time
and introduces a time dependence explicitly.
The two-particle BEC correlation function $R_2$ is then dependent on one two-particle variable,
$Q$, and on $a$, a parameter in the correlation between momentum space and space-time. 
The time-dependence is assumed to be given by an asymmetric L\'evy distribution, $H(\tau)$.
The parameter $a$ can be absorbed into an effective radius to obtain
\Eq{eq:asymlevR2},
which successfully describes BEC for both two- and three-jet events.
 
For two-jet events $a=1/\mt$ and the introduction of an effective radius is unnecessary.  The BEC
correlation
function is then a function not only of $Q$ but also of the transverse masses of the pions.  This
description of $R_2(Q,{\mt}_1,{\mt}_2)$ also successfully describes the data.
 
Nevertheless, the \taumodel\ description breaks down when confronted with data expressed in components
of $Q$.
The \taumodel\  predicts   that the only two-particle variable entering $R_2$ is $Q$.
There is no dependence on components of $Q$ separately.
This implies a spherical shape of the pion
region of homogeneity,
whereas previous analyses using static parametrizations have found a shape elongated along the event axis,
as evidenced by \Rside\ being smaller than \Rlong\ in the LCMS.  Assuming azimuthal symmetry about the event axis,
\Rside\ is the transverse radius of the region of homogeneity.
Accordingly, the \taumodel\ equations for $R_2$ have been modified \adhoc\/ to allow an elongation.
The elongation found
agrees with that found in previous analyses.
However, in the rest frame of the pair the radius parameter in the out direction is found to be approximately twice that in
the side direction. Thus the assumption of azimuthal symmetry about the event axis is found not to hold.
 
We note that elongation is expected~\cite{AR98a} for a hadronizing string in the Lund model,
where the transverse and longitudinal components of a particle's momentum arise from different mechanisms.
Perhaps the absence of azimuthal symmetry arises from gluon radiation, which is represented by kinks in the Lund string.
Therefore a possible improvement of the \taumodel\ could be to allow the longitudinal and transverse components
to have  different degrees of
correlation between a particle's momentum and its space-time production point. 

 
\begin{footnotesize}

\end{footnotesize}
 
 
\end{document}